\documentstyle[aps,epsf]{revtex}
\begin{document}
\author{Renato Vicente\footnote{rvicente@gibbs.if.usp.br} and Nestor Caticha
\footnote{nestor@gibbs.if.usp.br}}
\address{Instituto de F\'\i sica, Universidade de S\~ao Paulo\\
CP66318, CEP 05315-970, S\~ao Paulo, SP Brazil}
\title{Functional Optimisation of Online Algorithms in Multilayer Neural Networks. }
\date{May 97}
\maketitle

\begin{abstract}
{\small {\bf We study the online dynamics of learning in fully connected
soft committee machines in the student-teacher scenario. The locally optimal
modulation function, which determines the learning algorithm, is obtained
from a variational argument in such a manner as to maximise the average
generalisation error decay per example. Simulations results for the
resulting algorithm are presented for a few cases. The symmetric phase {\it %
plateaux} are found to be vastly reduced in comparison to those found when
online backpropagation algorithms are used. A discussion of the
implementation of these ideas as practical algorithms is given. }}
\end{abstract}



Learning how learning occurs in artificial systems has caught the attention
of the Statistical Mechanics community in the last decade. 
This interest was ignited by several reasons, among them, the invention of efficient learning-from-examples methods such as backpropagation, that permit learning in computationally complex machines, 
to the realisation that ideas from disordered systems, in particular spin glasses, could be applied to the study of attractor as well as feedforward neural networks and to the generalised interest in complex systems with rugged energy landscapes.

The main results from the Statistical Mechanics (see e.g.\cite
{seung,watkin,opper} ) approach have almost invariantly been obtained in the
thermodynamic limit and have benefited from the powerful techniques used to
calculate the averages over the disorder introduced by the random nature of
the examples.

Among several possible approaches to machine learning, online learning \cite
{amari} has been the subject of an intense research effort due to several
factors. In this scheme, examples are used only once, thereby avoiding the
need for expensive memory resources, typical of offline methods. This,
however, doesn't translate necessarily into poor performance since efficient
methods can be devised that have performance comparable to the memory based
ones. Furthermore, learning sequentially from single examples has a greater
biological flavor than offline processing. While efficiency, computational
economy and biological relevance may be the most relevant factors, the
theoretical possibility of rather complete analytical studies has also
played an important role. If each one of these factors is, by itself,
sufficiently important to make online learning an attractive scheme,
together they combine to give a most compelling argument for its thorough
study.

In this letter we present results of the optimisation of online supervised
learning in a model consisting of a fully connected multilayer feedforward
neural network, in what has become known as the student-teacher scenario.
The type of result we present here brings together two separate lines of
research that have been recently pursued by several groups.

The study of online backpropagation as put forward by Biehl and Schwarze 
\cite{biehlsch} and later developed in \cite{saasoll,riegbiehl} has
permitted the analytical understanding of several properties of the dynamics
of the learning process. The most striking feature being the existence of
learning {\it plateaux} or symmetric phases which signal learning stages
where the information available to the student and the form in which it is
used do not permit breaking the permutation symmetry among the hidden nodes.
Further learning eventually permits the escape from the neighbourhood of
these repulsive symmetric fixed points into the broken symmetry, specialized
phase. The onset of specialization and different methods to hasten it have
been dealt with by several authors \cite{saasoll2,biehlwo,saad1,amari2}.

The second line of research from which we draw is the variational study of
locally optimal online learning. This program deals with the determination
of lower bounds for the generalisation errors in different models in
controlled learning scenarios. The constructive nature of the variational
approach has permitted finding update rules that lead to student networks
with the optimal generalisation performance. The relation of this approach
to Bayesian methods has been discussed in \cite{KiCa} and in \cite{opper2}.

The variational method has been previously applied to machines with no
internal units \cite{caticha,BS93,BR94} or with hidden units but 
nonoverlapping receptive fields (RF) \cite{copelli,simonetti} and also in 
the case of unsupervised learning \cite{VDBR}. We will introduce the
variational method for feedforward machines with overlapping RF. The
differences stem from the fact that while in the former case the
generalization error is a monotonic decreasing function of the order
parameters (student-teacher overlaps), in the latter, the monotonicity is
lost, due to the appearance of crossed overlaps.

The main results here presented are the analysis of the locally optimized
online learning dynamics of a soft committee. We present results for
over-realisable and realisable cases. The striking reduction or complete
elimination of the {\it plateaux} in the learning curves witnesses the great
improvement achievable by concentrating in extracting the largest possible
amount of information from each example. Rapid escape from the plateaux can
be attributed to a fluctuation enhancing mechanism that stimulates
permutational symmetry breaking.

The aim of learning is to obtain a set of student weights $J_{ik}$ where $%
i(=1,...,N)$ indexes input layer units and $k(=1,...K)$ hidden nodes, in
such a manner that the student implements as closely as possible the map
represented by the teacher network defined by a set of weights $B_{in}$,
where $i(=1,...,N)$ labels the input layer unit and $n(=1,...M)$ the hidden
node. We use $n,m,...$ to label teacher branches and $%
j,k,...$ for the student branches. Call ${\bf B}
_n=(B_{1n},B_{2n},...,B_{Nn})$, ${\bf J}_k=(J_{1k},J_{2k},...,J_{Nk})$ the
weight branch vectors and $B_n$ and $J_k$ their respective lengths. We
define as usual the order parameters $R_{kn}={\bf J}_k\cdot {\bf B}_n$, $%
Q_{ij}={\bf J}_i\cdot {\bf J}_j$ and $M_{nm}={\bf B}_n\cdot {\bf B}_m$ which
will be taken $M_{nm}=\delta _{nm}$, for simplicity.

At each time step $\mu $, an example ${\bf S}^\mu $ is drawn from a known
distribution $P(S)$. We call $\Sigma _B^\mu $ and $\Sigma _J^\mu $ the
teacher and student outputs respectively. The internal fields are denoted by 
$y_n^\mu ={\bf B}_n\cdot{\bf S}^\mu $ and $x_k^\mu ={\bf J}_k\cdot{\bf S}
^\mu $. The available information is used in updating the student weights $%
J_{ik}$, 
\begin{equation}
J_{ik}(\mu +1)=J_{ik}(\mu )+\frac{F_k}NS_{i}^{\mu}
\end{equation}
This is not the most general update possible since a decay term, useful
in controlling the length of ${\bf J}_k$ can be used, we however will not
pursue this direction here. The central quantity in this theoretical
approach is the set of modulation functions ${\bf F}=(F_1, F_2, \dots, F_K)$%
. The following analysis will be done in the thermodynamic limit. For any
transfer function, the evolution of the order parameters is given by a set
of $(K^2+K)/2+KM$ first order differential equations. For fully connected
architectures we have: 
\begin{equation}
\frac{dR_{in}}{d\alpha }=\langle y_{n}F_i\rangle, \, \frac{dQ_{ij}}{d\alpha }
=\langle x_iF_j+x_jF_i+F_iF_j\rangle  \label{eq:din}
\end{equation}
where as usual, $\alpha =\mu /N$ measures the learning time. 
We now proceed, first to obtain the best ${\bf F}$, from a generalisation
point of view, and then to analyse the dynamical consequences that such a
choice will have.

A point of technical importance, which in no way restricts the validity of
the general properties of the results here discussed, concerns the choice of
an error function for the sigmoidal transfer function $g$ of the internal
units and a linear transfer function for the output unit, following \cite
{biehlsch}, since it permits better analytical tractability. Thus $%
\Sigma _B^\mu =\sum_{n=1,...,M}erf{(y_n^\mu /\sqrt{2})}$ and $\Sigma _J^\mu
=\sum_{k=1,..,K}erf{(x_k^\mu /\sqrt{2})}$.

For a fixed teacher, the student network will have a generalisation error $%
e_g({\bf J}_k)=\langle \frac 12(\Sigma _B^\mu -\Sigma _J^\mu )^2\rangle _{%
{\bf S}}$. In the thermodynamic limit, for a uniform distribution of
examples, the generalisation error can be written as a function of the
order parameters: 
\begin{eqnarray}
e_g &=&\frac 1\pi \sum_{i,j}\sin ^{(-1)}(\frac{Q_{ij}}{\sqrt{1+Q_{ii}}\sqrt{%
1+Q_{jj}}})  \nonumber \\
&-&\frac 2\pi \sum_{i,n}\sin ^{(-1)}(\frac{R_{in}}{\sqrt{2(1+Q_{ii})}})+%
\frac M6
\end{eqnarray}
Local optimisation
is obtained by maximizing the average generalisation error decay for each example in a given state $(R_{kn},Q_{ij})$. We thus look, following \cite{caticha}, at the extremes of the functional $\dot e_g[{\bf F}]=de_g[{\bf F}%
]/d\alpha $ that is, the modulation
function ${\bf F}$ which satisfies $\delta \dot e_g/\delta F_k=0$. The
solution has the general form 
\begin{equation}
{\bf F}={\bf H^{-1}G}\langle {\bf y}\rangle _{{\cal H}|{\cal V}}-{\bf x}
\label{eq:formula}
\end{equation}
where ${\bf H}$, the functional Hessian matrix and ${\bf G}$ are defined as $%
H_{ij}=\delta ^2\dot e_g/\delta F_i\delta F_j$ and $G_{kn}=-\partial
e_g/\partial R_{kn}$ and the conditional expectation is taken with respect
to the assumed examples' probability distribution $P(S)$. The symbols ${\cal %
H}$ and ${\cal V}$ stand for the set of ${\cal H}$idden or ${\cal V}$isible
information. It is interesting to note that (\ref{eq:formula}) holds for any
choice of transfer function or examples' distribution. For the particular
case of examples drawn independently from a uniform spherical distribution,
we have to solve integrals of the form: 
\begin{equation}
\int \prod_ndy_n\,P_{{\bf C}}({\bf x},{\bf y})y_m^\epsilon \delta (\Sigma _B-%
{\sum_n}erf(y_n^\mu /\sqrt{2}))
\end{equation}
where $\epsilon =0,1$ and $P_{{\bf C}}({\bf x},{\bf y})$ is a $(K+M)$
multivariate gaussian with correlation matrix 
\begin{equation}
{\bf C}=\left( 
\begin{array}{cc}
{\bf Q} & {\bf R} \\ 
{\bf R}^{{\bf t}} & {\bf M}
\end{array}
\right). 
\end{equation}

We now present results obtained by simulating an $N=5000$ system for the
cases $K=2$, $M=1$ and $K=M=2$. Further details will be presented elsewhere 
\cite{ViCa2}. In figures 1 and 2 we show the learning curves for these two
cases. Backpropagation results are included for comparison. Figure 3 shows
the evolution of $Q_{ik}$ for the $K=M=2$ case and suggests that the mechanism used to enhance fluctuations and break the permutation symmetry is to increase synaptic vector norms and stimulate
anti-correlated weights.

Whether this solution of the variational problem leads to a maximum 
generalisation or not will be governed by the functional Hessian 
matrix ${\bf H}$. Note that the dependence of the dynamics on the 
modulation function is only second order, therefore ${\bf H}$ is a 
function of the order parameters and not explicitly of the particular 
algorithm that led to that state of affairs. A negative eigenvalue of 
${\bf H}$ at a given point in the space of order parameters implies that 
at that point an optimal algorithm can not be analytically found.

The evolution of the eigenvalues for both cases is shown as insets in figures 1 and 2. In the space of algorithms, for both cases, at the beginning of the learning process these modulation functions represent saddle points rather than maxima. For the case

 $K=2$, $M=1$ this can be explained as follows. The best generalisation would be obtained by using a {\it correct }architecture, $K=M=1$, thus the optimal strategy is to trim the student into the correct architecture and then proceed with the optimized no

nlinear perceptron algorithm which could
then be obtained by the above variational method. This kind of modulation
function cannot be obtained analytically by searching for zero derivatives
in the space of algorithms of the $K=2$ student. The solution found by our
method does cut out one of the branches around $\alpha \approx 1$ and
turns itself into an effectively $K=1$ machine quite rapidly, avoiding the
long {\it plateau} of the backpropagation algorithm. 

The explanation for the initially negative eigenvalue of {\bf H} in the  
$K=M=2$ case is not different. The optimal strategy is within the space of 
students with a $K=2$ architecture and asymmetric inital conditions, and 
thus it will not be found by the variational approach. Before there is any 
information to hint that the permutation symmetry should be broken, it is 
more efficient locally to learn with a $K=1$ machine (with an output 
multiplied by 2). This is however not true after a while, since thus it 
will never escape the {\it plateau}. Since the escape is achieved by 
amplification of symmetry breaking fluctuations, learning initially with a 
nonlinear perceptron cannot be globally efficient, for it totally suppresses 
the desired effect of fluctuations.

A null eigenvalue of {\bf H} indicates the existence of a class of algorithms
with identical performance to that of eq.\ref{eq:formula}, this can be 
interpreted as a kind of {\it functional robustness}. An example of this 
appears for the case $K=M=2$, where the smallest eigenvalue stays very 
close to zero in the {\it plateau} state (see fig 2.). The significance of 
this is that due to functional robustness, the exact determination of the 
modulation function is not very critical for learning and eventually 
escaping the {\it plateau}.

Although our method permits the locally optimal extraction of information from an example, it does not assure
that the system will follow the best global trajectory in the space of order 
parameters. The global functional optimisation has been recently 
addressed in \cite{saad2}. They have shown the equivalence between local 
and global optimisations for the boolean perceptron and the better performance of the global approach in K=3, M=1 case. A thorough investigation
on how global and local optmisations are related is an important issue and 
remains to be done.

The effects of finite size $N$ have not been systematically investigated and
therefore the advantages of these methods, if any, over conventional
algorithms remains to be proved. Nevertheless, learning is easier in smaller
networks and a straightforward use of the modulation function in regimes where the
central limit theorem cannot yet be used leads to a successful learning
prescription as can be seen from simulating learning for the rather small
network with $N=15$, $K=M=2$ \cite{ViCa2}.

The main difficulties of using this approach to construct practical
algorithms concern the assumed knowledge of several unavailable quantities.
First of all the examples' probability distribution is needed in order to
calculate the integrals in equation (\ref{eq:din}). Then, the resulting
modulation function depends on unknown order parameters, such as $R_{in}$,
and worst, these order parameters are only self-averaging in the
thermodynamic limit. We first discuss rapidly the first two points.
Optimality is hard to define, several different possible criteria lead to
different results. Also, given a definition, such as the one we use here of
maximizing generalization, the optimal prescription will depend on the
amount of available information and on the environment where learning takes
place. Although we do not attempt to solve these problems here, a short
digression is in order. A parametric representation of $P(S,\Sigma
_B)\approx P_{{\bf w}}(S,\Sigma _B)$ permits introducing an extra set of $p$
differential equations for the online estimation of the distribution
parameters ${\bf w=}(w_1,w_2,...,w_p)$. Also the order parameters can be
analogously estimated online, as has been done in \cite{KiCa93}, even in the
case of time dependent or drifting rules.

How robust these ``optimal'' algorithms are in the absence or misestimation
of this information, as well as its response to learning in noisy
environments remains to be seen. The last issue has been addressed recently
in \cite{WUSP} for boolean machines. They found a large robustness to
noise-level-misestimation, as well as efficient online noise level
estimators which manage to steer the dynamics into an efficient learning
phase.

These comments about the need for extra knowledge to implement these methods
as algorithms can be seen as drawbacks for the variational program. We
rather think of them as calling our attention to the further work that has
to be done in order to obtain efficient adaptive practical algorithms, and
pointing out directions in which these objectives can be reached. Whatever
point of view is chosen, the validity of these results and their relation to
improving the generalisation ability remains.

 {\bf Acknowledgements }The authors thank O. Kinouchi
and M. Copelli for several useful discussions and M. Biehl, P. Riegler, S.
Solla and C. Van den Broeck for discussions during the early stages of this
work. This work received partial financial support from CNPq and Finep
(RECOPE).

\begin{figure}[tbp]
\epsfxsize = 0.6 \textwidth
\par
\begin{center}
\leavevmode
\epsfbox{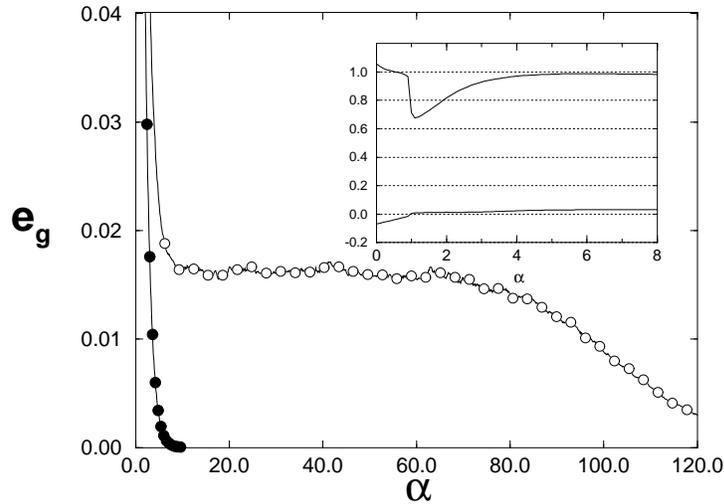}
\end{center}
\caption{Generalization error Learning Curves for the $K=2, M=1$ case
obtained by simulating a system of $N=5000$ with random initial conditions 
$Q_{11}\in[0,.5]$, $Q_{22}\in[0,1E-6]$ and $Q_{12}\approx 0$.
Black circles: optimized algorithm, white circles: conventional
backpropagation with learning rate $\eta =1.5$ . Inset: Eigenvalues of
the Hessian ${\bf H}$. There is a transient where the smallest eigenvalue
is negative, it then crosses rapidly into positive vales.}
\label{fig:1}
\end{figure}

\begin{figure}[tbp]
\epsfxsize = 0.6 \textwidth
\par
\begin{center}
\leavevmode
\epsfbox{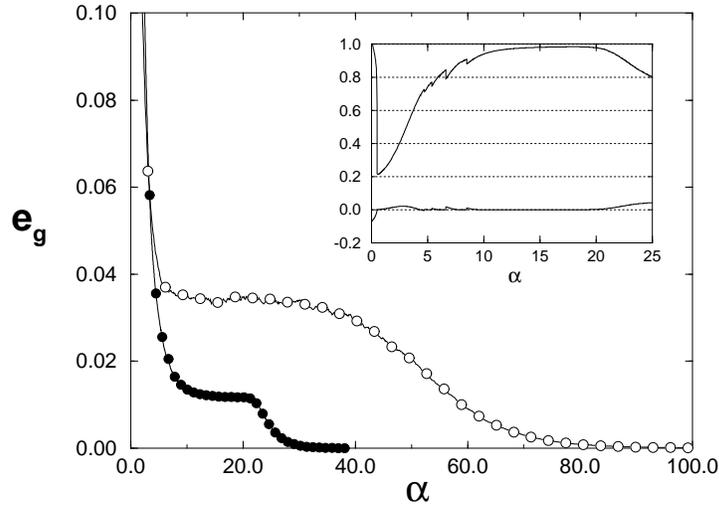}
\end{center}
\caption{Same as fig. 1 but for the $K=M=2$. Inset: Eigenvalues of the Hessian ${\bf H}$. Note that the smallest eigenvalue stays very close to zero in the {\it plateau}.}
\label{fig:2}
\end{figure}

\begin{figure}[tbp]
\epsfxsize = 0.6 \textwidth
\par
\begin{center}
\leavevmode
\epsfbox{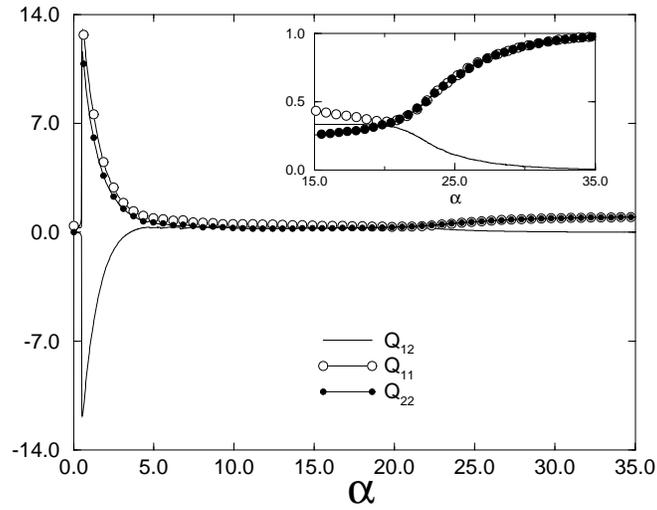}
\end{center}
\caption{Evolution of the overlaps. Note the anticorrelation that builds up during the transient. Inset: Details of the escape from the {\it plateau}.}
\label{fig:3}
\end{figure}

\end{document}